\documentclass[adp,fleqn]{w-art}
\usepackage{times}
\usepackage{w-thm}
\usepackage[]{graphicx}
\begin{document}

\DOIsuffix{theDOIsuffix}
\Volume{XX}
\Issue{x}
\Copyrightissue{x}
\Month{01}
\Year{2008}
\pagespan{1}{}
\Receiveddate{14 May 2008}
\Reviseddate{30 May 2008}
\Accepteddate{31 May 2008 by F.W.Hehl}
\Dateposted{5 July 2008}

\title[Electromagnetic energy-momentum]{Electromagnetic energy and momentum 
in moving media}

\author[Yuri N. Obukhov]{Yuri N. Obukhov 
  \footnote{E-mail:~\textsf{yo@thp.uni-koeln.de}} }
\address[]{Department of Theoretical Physics\\ 
Moscow State University\\ 117234 Moscow, Russia}

\begin{abstract}
  The problem of the electromagnetic energy-momentum tensor is among
  the oldest and the most controversial in macroscopic
  electrodynamics. In the center of the issue is a dispute about the
  Minkowski and the Abraham tensors for moving media. An overview of
  the current situation is presented. After putting the discussion
  into a general Lagrange-Noether framework, the Minkowski tensor is
  recovered as a canonical energy-momentum. It is shown that the
  balance equations of energy, momentum, and angular momentum are
  always satisfied for an open electromagnetic system despite the lack
  of the symmetry of the canonical tensor. On the other hand, although
  the Abraham tensor is not defined from first principles, one can
  formulate a general symmetrization prescription provided a timelike
  vector is available. We analyze in detail the variational model of a
  relativistic ideal fluid with isotropic electric and magnetic
  properties interacting with the electromagnetic field. The relation
  between the Minkowski energy-momentum tensor, the canonical
  energy-momentum of the medium and the Abraham tensor is clarified.
  It is demonstrated that the Abraham energy-momentum is relevant when
  the 4-velocity of matter is the only covariant variable that enters
  the constitutive tensor.
\end{abstract}
\maketitle 

\section{Introduction}

One century ago, in 1908, Hermann Minkowski \cite{minkowski} gave a
complete relativistically covariant formulation of classical
Maxwell-Lorentzian electrodynamics. In particular, he demonstrated how
the macroscopic physical processes in moving bodies and media can be
derived from the knowledge of the physics of the matter at rest.  An
important issue in phenomenological macroscopic electrodynamics of
moving media is the definition of the energy and momentum of the
electromagnetic field in matter. Rather surprisingly, this topic has
demonstrated a remarkable longevity, and the question of the
electromagnetic energy and momentum in matter was not settled until
now, despite some theoretical and experimental achievements.

It is not our aim to give a complete history of this issue. After 
Minkowski proposed \cite{minkowski} his version of the energy-momentum 
tensor, Abraham \cite{abraham1,abraham2,abraham3} entered the dispute 
with a different expression. The main formal difference between the 
Minkowski and the Abraham energy-momentum tensors is that the latter 
is symmetric while the former is not. Since then, the discussion was 
going on which of these two tensors is correct (along with some other 
attempts, among which it is worthwhile to mention the results of Einstein 
and Laub \cite{laub}, of de Groot and Suttorp \cite{degroot} and of 
Peierls \cite{Pei1,Pei2}).

Of the early studies, the work of D\"allenbach and Ishiwara
\cite{daelen1,daelen2,Ishiwara} deserves to be mentioned. The later
contributions of von Laue, Beck, Haus \cite{laue1,laue2,beck,haus} and
many others
\cite{Gram,Tis,Ott,Sch,Marx,Nov,Nag1,Nag2,Sho,Bal,Jauch1,Jauch2,Jauch3}
and
\cite{skob1,skob2,ginz1,ginz2,ginz3,Gy,Wong,Lai1,Lai2,Arn,Agu,Tang,Fu}
brought more arguments and more confusion into the discussion.
Eventually, it became clear that the study of the electromagnetic
field alone is insufficient, and the idea was put forward that all
possible tensors are correct, depending on the way how field and
matter are separated into two subsystems, see for example
\cite{Kranys1,Kranys2,Lorrain,Maugin} and
\cite{Israel1,Israel2,Israel3}. More recent discussions were devoted
to the analysis of this separation question
\cite{Brevik1,Brevik2,Brevik3,Brevik4,Brevik5,Lai3,gordon}, along with
\cite{tier1,tier2,Nelson,Loudon1,Loudon2,Mansur,Chiao} and
\cite{antoci1,antoci2,antoci3}.

{}From a theoretical point of view, the most important developments
should perhaps be attributed to Schmut\-zer and Schr\"oder
\cite{schmutzer1,schmutzer2,schmutzer3,Schroeder} who stressed the
relation of the Minkowski and the canonical energy-momentum tensor, to
Poincelot \cite{poin1,poin2,poin3,poin4} for underlining the
universality of the Lorentz force, and to Penfield, Haus, and Mikura
\cite{Penfield1,Penfield2,Penfield3,Mikura} for coming up with the
first working models for polarizable and magnetizable matter in the
framework of the Lagrange variational approach. In our paper, these
points will be put into the center of the discussion, following our
earlier studies \cite{emt,emtme}. Some relevant results were also
presented in the recent interesting papers of Dereli et al.\
\cite{Dereli1,Dereli2}.

It is clearly impossible to mention even briefly all the names and
to discuss all the contributions to the issue of the electromagnetic
energy and momentum in a short paper. Moreover, since we will mainly
pay attention to the theoretical questions here, the experimental 
results in this area will be not be properly discussed. In order
to get more information, the readers should consult the thorough 
reviews \cite{degroot,Pauli,LL60,Moller,Fano,Novak,robin,brevik}, see
also the most recent one in \cite{newrmp}. 

Our basic notations and conventions follow the book \cite{HO03}. In
particular, $\varepsilon_0, \mu_0$ are the electric and the magnetic
constant (earlier called vacuum permittivity and vacuum permeability).
It is worthwhile to stress that we do not choose any special system of
physical units. In other words, the presence of $\varepsilon_0$ and
$\mu_0$ in our formulas does not mean that the international system SI
is used, their values are different for SI, Lorentz-Heaviside, and
rational systems, and they are fixed only after the specific unit system 
is chosen. More on physical dimensions and units can be found in 
\cite{okun}. We do not discuss gravitational effects, the spacetime 
is flat. The Minkowski metric is $g_{ij}={\rm diag}(c^2,-1,-1,-1)$.  
The totally antisymmetric Levi-Civita tensor $\eta_{ijkl}$ is defined
by $\eta_{0123} = \sqrt{-g}$, with $g = {\rm det}g_{ij}$. Latin indices 
from the middle of the alphabet label the spacetime components, $i,j,k,
\dots=0,1,2,3$, whereas those from the beginning of the alphabet refer 
to 3-space: $a,b,c,\dots=1,2,3$.

\section{Preliminaries: macroscopic electrodynamics}\label{maxwell}

Quite generally, the Lagrangian of the {\it electromagnetic field in matter}
can be written as 
\begin{equation}
L^{\rm e} = -\,{\frac 18}\,\chi^{ijkl}\,F_{ij}F_{kl}.\label{Le}
\end{equation}
Here $F_{ij} = \partial_iA_j - \partial_jA_i$ is the electromagnetic field 
strength constructed from the potential $A_i$, and the tensor density 
$\chi^{ijkl}$ describes the electric and magnetic properties of matter. 
This quantity may also depend on $F_{ij}$, thereby taking into account 
possible nonlinear electromagnetic effects, but in the first approximation 
$\chi^{ijkl}$ is a function only of the matter variables that describe 
the state of the medium. We will restrict ourselves to {\it linear 
electrodynamics}. 

As usual, we define the {\it electromagnetic excitation} tensor density by 
\begin{equation}
H^{ij} = -\,2{\frac {\partial L^{\rm e}}{\partial F_{ij}}}.\label{Hij}
\end{equation}
For the theory (\ref{Le}) we then find the linear constitutive relation
\cite{Tamm1,Tamm2,Tamm3,Post}
\begin{equation}
H^{ij} = {\frac 12}\,\chi^{ijkl}\,F_{kl}.\label{HF}
\end{equation}
Accordingly, we then have $L^{\rm e} = -\,{\frac 14}\,H^{kl}F_{kl}$. 

In the conventional way, the components of the tensors $F_{ij}$ and
$H^{ij}$ describe the electric and magnetic fields $({\bf E}, {\bf
  B})$ and the electric and magnetic excitations $({\bf D}, {\bf H})$
(other names of which are ``electric displacement" and ``magnetic
field intensity"):
\begin{equation}
F_{ij} = \left(\begin{array}{cccc} 0 & -E_1 & - E_2 & -E_3 \\
E_1 & 0 & B^3 & - B^2 \\ E_2 & -B^3 & 0 & B^1 \\ E_3 & B^2 & - B^1 & 0
\end{array}\right),\qquad
H^{ij} = \left(\begin{array}{cccc} 0 & D^1 & D^2 & D^3 \\
-D^1 & 0 & H_3 & - H_2 \\ -D^2 & -H_3 & 0 & H_1 \\ -D^3 & H_2 & - H_1 & 0
\end{array}\right).\label{FHij}
\end{equation}
It is worthwhile to mention that it was Minkowski \cite{minkowski} who
introduced the notion of a bivector (which he called ``Raum-Zeit-Vektor 
II$^{\rm ter}$ Art", i.e., a second rank tensor in the modern terminology) 
and wrote for the first time the Maxwell electrodynamics in the explicitly 
covariant four-dimensional tensor form. 

In vacuum, the electromagnetic action reads $I = -\,{\frac 14}\int \lambda_0
g^{ik}g^{jl}F_{ij}F_{kl}\,\sqrt{-g}d^4x$, where $\lambda_0 = \sqrt{
\varepsilon_0/\mu_0}$. As a result, in vacuum we find for the constitutive
tensor density
\begin{equation}
{\stackrel 0 \chi}{}^{ijkl} = \lambda_0\sqrt{-g}\left(g^{ik}g^{jl} 
- g^{il}g^{jk}\right).\label{chi0}
\end{equation}
Consequently, the linear constitutive relation (\ref{HF}) in vacuum 
in Cartesian coordinates reads (note that $\sqrt{-g} = c$, $\lambda_0c 
= \mu_0^{-1}$, and $\lambda_0/c = \varepsilon_0$) 
\begin{equation}\label{HF0}
H^{ij} = {\frac 12}\,{\stackrel 0 \chi}{}^{ijkl}\,F_{kl} = F^{ij}/\mu_0, 
\end{equation}
or, equivalently, in 3-dimensional form
\begin{equation}\label{HF0v}
{\bf D} = \varepsilon_0\,{\bf E},\qquad {\bf H} = {\frac 1{\mu_0}}\,{\bf B}.
\end{equation}

In macroscopic electrodynamics\footnote{See Hirst \cite{hirst} for the
  definition and a discussion of magnetization and polarization in the
  microscopic theory.}, a polarizable and magnetizable medium is
characterized by the nontrivial polarization ${\bf P}$ and
magnetization ${\bf M}$. They arise from the bound charge and current
densities $({\stackrel p \rho}, {\stackrel p {\bf j}})$ via the
standard relations
\begin{equation}
{\stackrel p \rho} = -\,{\bf \nabla}\cdot{\bf P},\qquad {\stackrel p {\bf j}}
= \dot{\bf P} + {\bf \nabla}\times {\bf M}.\label{PM}
\end{equation}
The vacuum constitutive law (\ref{HF0v}) in a medium is changed 
\cite{Kovetz} into
\begin{equation}\label{HF1v}
{\bf D} = \varepsilon_0\,{\bf E} + {\bf P},\qquad {\bf H} = {\frac 1{\mu_0}}
\,{\bf B} - {\bf M}.
\end{equation}
Analogously to the electric and magnetic fields, the polarization and 
magnetization are not 3-vectors, but constitute the components of the 
4-dimensional polarization tensor of second rank,
\begin{equation}
M^{ij} = \left(\begin{array}{cccc} 0 & P^1 & P^2 & P^3 \\
-P^1 & 0 & -M_3 & M_2 \\ -P^2 & M_3 & 0 & -M_1 \\ -P^3 & -M_2 & M_1 & 0
\end{array}\right).\label{Mij}
\end{equation}
Correspondingly, the constitutive relation (\ref{HF1v}) can be recast
into
\begin{equation}
H^{ij} = {\frac 1 {\mu_0}}\,F^{ij} + M^{ij},\label{HF1}
\end{equation}
and, introducing the 4-vector of the {\it bound current} ${\stackrel p
  J}{}^i = ({\stackrel p \rho},\,{\stackrel p {\bf j}})$, equation
(\ref{PM}) can be rewritten as
\begin{equation}
{\stackrel p J}{}^i = -\,\partial_jM^{ij}.\label{JMP}
\end{equation}

Combining (\ref{HF}) and (\ref{HF0}), we can put the constitutive
relation in an alternative equivalent form:
\begin{equation}
M^{ij} = {\frac 12}\,\xi^{ijkl}\,F_{kl},\qquad \xi^{ijkl} := 
\chi^{ijkl} - {\stackrel 0 \chi}{}^{ijkl}.\label{MF}
\end{equation}
In general, the {\it susceptibility tensor density} $\xi^{ijkl}$ describes
also the magnetoelectric effects when the medium is magnetized in an 
electric field and/or polarized in a magnetic field. 
  
\section{Abrahamization of an energy-momentum tensor}\label{abr}

Contrary to a general belief (and perhaps somewhat surprisingly) there
does not exist any derivation of the Abraham energy-momentum tensor
{\it from first principles}. In this section, we give a formal
definition of what can be called the ``abrahamization" prescription
that can be applied to an arbitrary energy-momentum tensor.

More precisely, let us assume that we can associate with a given
physical system a second rank tensor $t_k{}^i$, the components of
which are interpreted as follows: $t_0{}^0 = w$ is a density of energy
of the system, $t_0{}^a = s^a$ (with $a = 1,2,3$) is the energy flux
density (``Poynting vector"), $t_a{}^0 = - p_a$ is the momentum
density, and $t_a{}^b$ is the stress tensor. We do not assume that
this tensor is conserved, and the divergence $\partial_i t_k{}^i =
f_k$ describes, in general, the balance equation of the system. For $k
= 0$, this yields the familiar energy balance equation $\dot{w} +
{\bf\nabla} \cdot{\bf s} = f_0$. Integrating over a three-dimensional
volume $V$, we find in a usual way that the change in time of the
total energy $\int_V w$ combined with the energy flux through the
boundary, $\int_{\partial V} s$, is equal to the total power
production of the system $\int_V f_0$. Similarly, for $k = a = 1,2,3$
we recover the balance equation $-\dot{p}_a + \partial_b\,t_a{}^b =
f_a$ of the momentum of the system under the action of the force $f_a$. The
4-vector force density vanishes, $f_k = 0$, for a closed physical
system that does not interact with other systems.

Let us raise the first index, $t^{ij} = g^{ik}t_k{}^j$. The resulting
covariant tensor is not symmetric in general: $t^{ij}\neq t^{ji}$. In
particular, this means that the momentum density is not equal to the
energy flux density, ${\bf p} \neq {\bf s}/c^2$. There is nothing
unphysical about this fact which may be explained, for example, by the
nontrivial intrinsic angular momentum (spin) of the elements of the
system \cite{HehlRoMP}. A well known symmetrization prescription of
Belinfante and Rosenfeld \cite{Belinfante1,Belinfante2,Rosenfeld}
allows one to construct from the original $t_k{}^i$ and from the spin
density tensor another energy-momentum tensor ${\stackrel {B}
  t}{}_k{}^i$ which is symmetric, ${\stackrel {B} t}{}^{ij} =
{\stackrel {B} t}{}^{ji}$, and has the same divergence $\partial_i
t_k{}^k = \partial_i{\stackrel {B} t}{}_k{}^i$.

A different symmetrization procedure is possible for the case when a
timelike vector field $u^i$ is available, with $u^2 = u_iu^i > 0$.
Physically, $u^i$ can be interpreted as a 4-velocity of the system
with respect to the inertial reference frame. With the help of this
field, we can introduce the transversal and longitudinal projectors
${\stackrel \bot P}{}_j^i = \delta_j^i - u^iu_j/u^2$ and ${\stackrel
  {||} P}{}_j^i = u^iu_j/u^2$. Let us denote the antisymmetric part of
the original energy-momentum tensor by $a^{ij} := t^{[ij]} = (t^{ij} -
t^{ji})/2$. Then we define
\begin{equation}
{\stackrel {A} t}{}_k{}^i := t_k{}^i - {\stackrel \bot P}{}_k^ja_j{}^i 
- {\stackrel {||} P}{}_j^ia_k{}^j.\label{tAdef}
\end{equation}
By construction, this object is symmetric,
\begin{equation}
{\stackrel {A} t}{}^{ij} = t^{(ij)} - {\frac 2{u^2}}\,u^{(i}u_k a^{j)k},\label{tA}
\end{equation}
and has the following crucial property: For the system {\it at rest},
$u^i = \delta^i_0$, the time-time and space-space components coincide
with the original symmetrized ones, ${\stackrel {A} t}{}_0{}^0 =
t_0{}^0 = w$ and ${\stackrel {A} t} {}^{ab} = t^{(ab)}$, whereas the
off-diagonal (time-space and space-time) components read
\begin{equation}
{\stackrel {A} t}{}^{a0} = t^{a0} - 2a^{a0} = t^{0a},\qquad
{\stackrel {A} t}{}^{0a} = t^{0a}.\label{tAs}
\end{equation}
In other words, the energy flux density (``Poynting vector") remains
the same as before, ${\stackrel {A} s}{}^a = {\stackrel {A} t}{}_0{}^a
= t_0{}^a = s^a$, but the momentum density $p^a$ is replaced with
${\stackrel {A} p}{}^a = {\stackrel {A} t}{}^{a0} = t^{0a} = s^a/c^2$.

The Belinfante-Rosenfeld procedure and the {\it abrahamization}
(\ref{tAdef}) of the energy-momentum tensor both produce symmetric
tensors from a given original asymmetric energy-momentum $t_k{}^i$.
However, these two symmetrization schemes are distinct in the
following significant point. For the Belinfante-Rosenfeld approach,
the balance equation remains untouched because the divergence
$\partial_i{\stackrel {B} t}{}_k {}^i = \partial_i t_k{}^i$ is
preserved. In contrast, the divergence $\partial_i {\stackrel {A}
  t}{}_k{}^i \neq \partial_i t_k{}^i$. Let us find the difference of
the two force densities, ${\stackrel {A} f}{}_k = \partial_i{\stackrel
  {A} t}{}_k {}^i$ and $f_k = \partial_i t_k{}^i$. From (\ref{tAdef}),
\begin{equation}
{\stackrel {A} f}{}_k = f_k - \partial_i({\stackrel \bot P}{}_k^ja_j{}^i)
- \partial_i({\stackrel {||} P}{}_j^ia_k{}^j).\label{diff1}
\end{equation}
In the rest frame, $u^i = \delta^i_0$, we find ${\stackrel {A} f}{}_0 = f_0$
and 
\begin{equation} {\stackrel {A} {\bf f}}{} = {\bf f} + {\frac
    {\partial}{\partial t}}\left({\bf p} - {\frac {\bf s}{c^2}}\right)
  + {\bf \nabla}\times {\bf a}.\label{diff2}
\end{equation}
We use the boldface notation for the spatial 3-dimensional objects. In
particular, ${\bf f} = \{f_a\}$, ${\bf p} = \{p_a\}$, ${\bf s} =
\{s_a\}$, whereas the antisymmetric part of the stress tensor gives
rise to ${\bf a} = \{{\frac 12} \epsilon_{abc}a^{bc}\}$.

As such, the definition (\ref{tAdef}) looks rather artificial. To the
best of my knowledge, there does not exist any derivation of
(\ref{tAdef}) from first principles, such as from the Lagrange-Noether
machinery, for example. One merely demands the symmetry of the
energy-momentum tensor under the condition that the energy density and
the energy flux density in the rest frame remain the same. The form of
the resulting ${\stackrel {A} t}{}_k{}^i$ is then fixed by
(\ref{tAdef}) which leads to the additional ``Abraham force" terms
(\ref{diff2}) in the balance equations.

\section{Canonical energy-momentum for open systems}\label{open}

It is sometimes claimed that the symmetry is a fundamental property of
the energy-momentum tensor that is related to the conservation of the
angular momentum of the system. In order to clarify this point, let us
recall the relevant facts of the Lagrange and Noether theory. Suppose
we have a system of the fields denoted collectively by $\Phi^A$. Its
dynamics is described by the action integral $I = \int L\,d^4 x$,
where the Lagrangian (density) is a function of the fields and their
derivatives, $L = L(\Phi^A,\partial_i\Phi^A)$, and $x^i = (t,x^a)$ are
the (local) spacetime coordinates.

When the action $I$ is invariant under the {\it spacetime
  translations}, one finds the conservation law\footnote{We omit the
  derivations of these results which are well known since the work of
  Noether \cite{Noether}; for the detailed discussion see Sec.~19 of
  \cite{Corson}, for example.}
\begin{equation}\label{dt}
  \partial_i\,\Sigma_k{}^i = -\,{\frac {\delta L}{\delta\Phi^A}}\,
\partial_k\Phi^A.
\end{equation}
Here the {\it canonical energy-momentum tensor} is defined by
\begin{equation}
\Sigma_k{}^i = {\frac {\partial L}{\partial(\partial_i\Phi^A)}}\partial_k\Phi^A
- L\,\delta^i_k,\label{emt}
\end{equation}
and we use standard notation for the the variational derivative,
\begin{equation}
{\frac {\delta L}{\delta\Phi^A}} := {\frac {\partial L}{\partial\Phi^A}}
- \partial_i\left({\frac {\partial L}{\partial(\partial_i\Phi^A)}}\right).
\end{equation}

Similarly, assuming that action is invariant under the {\it Lorentz
  group}, that acts on the coordinates and fields, we find
\begin{equation}
\Sigma_{[jk]} + \partial_i\,S_{jk}{}^i = -\,{\frac {\delta L}{\delta\Phi^A}}
\,(\rho_{jk})^A_B\,\Phi^B.\label{dS}
\end{equation}
Here $(\rho^j{}_i)^A_B$ are the Lorentz generators for the fields $\Phi^A$, and
the {\it spin current density} is introduced by 
\begin{equation}
S_{jk}{}^i = {\frac {\partial L}{\partial(\partial_i\Phi^A)}}
\,(\rho_{jk})^A_B\,\Phi^B.\label{spin}
\end{equation}
Recall that the Lorentz generators are skew-symmetric, $(\rho_{jk})^A_B
= -\,(\rho_{kj})^A_B$. Hence the spin is skew-symmetric too, $S_{jk}{}^i
= -\,S_{kj}{}^i$. 

When the physical system is closed in the sense that it does not
interact with other systems, then its state is completely described by
the variables $\Phi^A$ and their dynamics in time and space. The
latter is determined by the field equations, ${\delta
  L}/{\delta\Phi^A} = 0$. Using this in the conservation laws
(\ref{dt}) and (\ref{dS}), we obtain $\partial_i\,\Sigma_k{}^i = 0$
and $\Sigma_{[jk]} + \partial_i\,S_{jk}{}^i = 0$. Accordingly, we find
that the canonical energy-momentum tensor is symmetric when the spin
is trivial, $S_{jk}{}^i = 0$. Otherwise, we can use the
Belinfante-Rosenfeld \cite{Belinfante1,Belinfante2,Rosenfeld}
procedure to derive the symmetric tensor $\sigma_k{}^i = \Sigma_k{}^i
-\partial_j\left(S^{ij}{}_k + S_k{}^{ji} - S_k{}^{ij}\right)$. By
construction, we have $\partial_i\,\sigma_k{}^i =
\partial_i\,\Sigma_k{}^i$, and $\sigma_{[ik]} = 0$.

However, the above is true for {\it closed systems only}. For open
systems that interact with other systems, we {\it cannot put} ${\delta
  L}/{ \delta\Phi^A} = 0$ and we have to keep the corresponding terms
on the right hand sides of (\ref{dt}) and (\ref{dS}). These terms
describe forces and torques which result from the interaction of the
systems. Even when the spin is absent, the canonical energy-momentum
of an open physical system is, in general, necessarily asymmetric in
order to maintain the balance of the nontrivial torque present on the
right-hand side of (\ref{dS}). We will see below how this works for
the electromagnetic field interacting with polarizable and
magnetizable matter.

\section{Energy-momentum tensors in electrodynamics}\label{EM}

After the preparations done in the previous three sections, we are now
in a position to start the main discussion. Here we consider the 
construction of the canonical and symmetric energy-momentum tensors in 
the {\it macroscopic electrodynamics} of media. The corresponding 
Lagrangian approach was formulated in Sec.~\ref{maxwell}. Our starting
point is the action $I^{\rm e} = \int L^{\rm e} d^4x$.

Without loosing generality, we can choose the collective field of the
system as $\Phi^A = (A_i, \chi^{ijkl})$. The set of all fields is thus
naturally divided into the two sectors, an electromagnetic and a
material one. It is not necessary to specify how the tensor density
$\chi^{ijkl}$ depends on the more fundamental material variables (we
can always do this at a later stage), but instead it is convenient to
treat $\chi^{ijkl}$ itself as a generalized material variable.

\subsection{Canonical energy-momentum tensor}

The Lagrangian (\ref{Le}) contains only derivatives of the
electromagnetic potential $A_i$ but not of $\chi^{ijkl}$, and thus we 
easily derive from (\ref{emt}) and (\ref{Le}) the canonical energy-momentum 
tensor of the system:
\begin{equation}
{\stackrel {\rm e} \Sigma}{}_k{}^i = -\,H^{ij}F_{kj} - L^{\rm e}
\,\delta_k^i = -\,H^{ij}F_{kj} + {\frac 14}H^{jl}F_{jl}\,\delta_k^i.\label{emtE}
\end{equation}
This is the well known {\it Minkowski energy-momentum}. Since the
system is obviously {\it open}, the canonical tensor is not conserved.
This becomes clear when we calculate the variational derivatives that
enter the right-hand side of (\ref{dt}). Indeed, we find explicitly
\begin{equation}\label{LL}
\Lambda^i := {\frac {\delta L^{\rm e}}{\delta A_i}} = - \partial_jH^{ij},\qquad
{\frac {\delta L^{\rm e}}{\delta \chi^{ijkl}}} = -\,{\frac 18}\,F_{ij}F_{kl}.
\end{equation}
Both variational derivatives are clearly nontrivial. We should take
into account that this (electromagnetic field) system interacts with
the matter, and the total Lagrangian is thus the sum $L^{\rm tot} =
L^{\rm e} + L^{\rm m}$. After introducing the electric current $J^i =
{\delta L^{\rm m}}/{\delta A_i}$, one then derives the Maxwell field
equation
\begin{equation}
{\frac {\delta L^{\rm tot}}{\delta A_i}} = \Lambda^i + J^i 
= - \partial_jH^{ij} + J^i = 0.\label{Max}
\end{equation}
This shows that $\Lambda^i$ vanishes only in the absence of electric
charge and current densities of matter, but in general $\Lambda^i =
-J^i\neq 0$.  Substituting (\ref{LL}) into (\ref{dt}), we derive the
energy-momentum balance equation for linear Maxwellian
electrodynamics as
\begin{equation}
\partial_i{\stackrel {\rm e} \Sigma}{}_k{}^i = F_{ki}J^i + X_k,\qquad
X_k = {\frac 18}\,F_{ij}F_{mn}\,\partial_k\chi^{ijmn}.\label{dtM}
\end{equation}
The first term on the right-hand side is the familiar Lorentz force,
whereas the second term requires some analysis. At first, we notice
that
\begin{equation}\label{X1}
X_k = {\frac 14}\left(F_{ij}\partial_k H^{ij} - H^{ij}\partial_k F_{ij}\right) 
= {\frac 14}\left(F_{ij}\partial_k M^{ij} - M^{ij}\partial_k F_{ij}\right),
\end{equation}
where we used (\ref{HF1}). Now, after some straightforward algebra, we
can bring this expression into the form
\begin{equation}\label{X2}
X_k = -\,\partial_i {\stackrel {\rm p} \Sigma}{}_k{}^i - F_{ki}\partial_jM^{ij},
\end{equation}
after introducing the {\it polarizational energy-momentum} tensor
\begin{equation}
{\stackrel {\rm p} \Sigma}{}_k{}^i := M^{ij}F_{kj} 
- {\frac 14}M^{jl}F_{jl}\,\delta_k^i.\label{emtP}
\end{equation}
The final step is to recall (\ref{JMP}) and to substitute (\ref{X2}) into 
(\ref{dtM}). After rearranging the terms, the result reads:
\begin{equation}
\partial_i\left({\stackrel {\rm e} \Sigma}{}_k{}^i + {\stackrel {\rm p} \Sigma}
{}_k{}^i\right) = F_{ki}\left(J^i + {\stackrel {\rm p} J}{}^i\right).\label{dtT}
\end{equation}

The form of this equation suggests to define the {\it total electromagnetic 
energy-momentum tensor} and the {\it total electric current} as the sums
\begin{equation}
{\stackrel {\rm tot} \Sigma}{}_k{}^i := {\stackrel {\rm e} \Sigma}{}_k{}^i 
+ {\stackrel {\rm p} \Sigma}{}_k{}^i,\qquad {\stackrel {\rm tot} J}{}^i 
:= J^i + {\stackrel {\rm p} J}{}^i.\label{defTJ}
\end{equation}
The physical interpretation is clear: $J^i$ is the current density of
the {\it free} charges, and ${\stackrel {\rm p} J}{}^i$ is the
polarizational current density of the {\it bound} charges. The
corresponding energy and momentum, associated with the free and bound
charges, are described by ${\stackrel {\rm e} \Sigma}{}_k{}^i$ and
${\stackrel {\rm p} \Sigma}{}_k{}^i$, respectively. As a result, the
balance equation (\ref{dtT}) is recast into
\begin{equation}\label{dT}
\partial_i{\stackrel {\rm tot} \Sigma}{}_k{}^i = F_{ki}\,{\stackrel {\rm tot} J}{}^i,
\end{equation}
with the Lorentz force on the right-hand side that acts on all types
of charges (free and bound) present in the medium.

Combining (\ref{emtE}) and (\ref{emtP}), and using the constitutive relation
(\ref{HF1}), we derive the explicit form of the total energy-momentum:
\begin{equation}
{\stackrel {\rm tot} \Sigma}{}_k{}^i = (-H^{ij} + M^{ij})F_{kj} 
+ {\frac 14}(H^{jl} - M^{jl})F_{jl}\,\delta_k^i = {\frac 1 {\mu_0}}\left(
-\,F^{ij}F_{kj} + {\frac 14}F^{jl}F_{jl}\,\delta_k^i\right).\label{emtT}
\end{equation}
This energy-momentum tensor was discussed by Poincelot
\cite{poin1,poin2,poin3,poin4} and more recently by us \cite{emt}.

Quite remarkably, we discover that the form of the total
electromagnetic energy-momentum tensor in media is precisely the same
as in vacuum. It is worthwhile to stress that the final result
(\ref{dT}) is a direct consequence of the fact that the
electromagnetic field is an {\it open system} (\ref{Le}). The energy
and momentum of the electromagnetic field are described by the {\it
  canonical (=Minkowski)} tensor (\ref{emtE}), but since the system is
open, the balance equation (\ref{dtM}) contains nontrivial force terms
that arise from the interaction of the field with matter. A careful
evaluation of the force $X_k$ then eventually brings the balance
equation into the final form (\ref{dT}). This general result is the
best what can be done without entering into the question about the
structure and dynamics of the medium.  In the next section we will
discuss a specific model of matter and demonstrate how the additional
knowledge of the physical nature of a medium can provide a different
computation of the force term $X_k$.

\subsection{Balance of the angular momentum}

The canonical energy-momentum (\ref{emtE}) is not symmetric. However,
this does not mean that there is a problem with the angular momentum
conservation. We simply have to recall once again that the system is
open, and hence the right-hand side of the angular momentum balance
equation (\ref{dS}) does not vanish, because the variational
derivatives (\ref{LL}) are nontrivial.

The Lorentz generators for the material variable $\chi^{ijkl}$ read
\begin{equation}
(\rho^j{}_k){}^{mnpq}_{m'n'p'q'} = \delta^{[j}_{m'}\delta^m_{k]}
\delta^n_{n'}\delta^p_{p'}\delta^q_{q'} + \delta^m_{m'}\delta^{[j}_{n'}
\delta^n_{k]}\delta^p_{p'}\delta^q_{q'} + \delta^m_{m'}\delta^n_{n'}
\delta^{[j}_{p'}\delta^p_{k]}\delta^q_{q'} + \delta^m_{m'}\delta^n_{n'}
\delta^p_{p'}\delta^{[j}_{q'}\delta^q_{k]}.
\end{equation}
Using (\ref{LL}) we then immediately find 
\begin{equation}
-\,{\frac {\delta L^{\rm e}}{\delta \chi^{mnpq}}}(\rho_{jk})^{mnpq}_{m'n'p'q'}
\chi^{m'n'p'q'} = {\frac 18}F_{[k|n}F_{pq|}\,4\chi_{j]}{}^{npq} = F_{[k|n|}
H_{j]}{}^n.
\end{equation}
We used (\ref{HF}) here. On the other hand, the antisymmetric part of
the canonical energy-momentum (\ref{emtE}) is straightforwardly found
to be
\begin{equation}\label{emtMa}
{\stackrel {\rm e} \Sigma}{}_{[jk]} = -\,H_{[k}{}^nF_{j]n} = F_{[k|n|}H_{j]}{}^n.
\end{equation}
We thus demonstrated that the balance equation (\ref{dS}) of the
angular momentum,
\begin{equation}
{\stackrel {\rm e} \Sigma}{}_{[jk]} \equiv -\,{\frac {\delta L^{\rm e}}{\delta 
\chi^{mnpq}}}(\rho_{jk})^{mnpq}_{m'n'p'q'}\chi^{m'n'p'q'},\label{angbal}
\end{equation}
is satisfied {\it identically} for the Minkowski canonical
energy-momentum. In physical terms, the skew part of the
energy-momentum tensor is perfectly balanced by the torque which is
present on the right-hand side because of the interaction of the field
with matter.

There is a somewhat subtle point which we silently avoided in our
previous discussion. However, for completeness we have to mention it
now. Namely, it is well known that there exist two possible choices of
the electromagnetic variable: {\it (i)} We can take the 1-form $A$ as
fundamental variable, and construct the corresponding Lagrangian as a
function of the field strength 2-form $F = dA$. This is our choice
which we assumed in the book \cite{HO03} and also here. The potential
1-form $A$ is an invariant object independent of local coordinates and
frames, and it does not transform under Lorentz rotations. The
canonical energy-momentum 3-form then turns out to be explicitly
gauge-invariant, given in tensor language by the expression
(\ref{emtE}), while the spin density formally vanishes. {\it (ii)} The
second choice is to take the components $A_i$ of a covector. They
transform (with a usual tensorial law) under the Lorentz group, and
the corresponding generators read $(\rho^j{}_k)^m_n = -
\delta^{[j}_n\delta^m_{k]}$. For such a choice of the field variable,
the energy-momentum tensor is given by the gauge noninvariant
expression which is obtained from (\ref{emtE}) by the shift
${\stackrel {\rm e} \Sigma}{}_k{}^i\rightarrow {\stackrel {\rm e}
  \Sigma} {}_k{}^i + \Delta\Sigma_k{}^i$, with an additional term
$\Delta\Sigma_k{}^i = -H^{ij}\partial_jA_k$. Since
$(\rho^j{}_k)^m_nA_m = -\,\delta^{[j}_nA_{k]}$, we find from
(\ref{spin}) the spin tensor $S_{jk}{}^i = H^i{}_{[j}A_{k]}$. Its
divergence thus contributes to the left-hand side of (\ref{dS}) which
now reads
\begin{equation}
\Delta\Sigma_{[jk]} + \partial_iS_{jk}{}^i = -\,H_{[k}{}^i\partial_{|i|}A_{j]}
+ \partial_i(H^i{}_{[j}A_{k]}) = (\partial_iH^i{}_{[j})A_{k]}.
\end{equation}
But at the same time, a new term appears on the right-hand side of (\ref{dS}):
\begin{equation}
-\,{\frac {\delta L^{\rm e}}{\delta A_n}}(\rho_{jk})^m_nA_m = \Lambda_{[j}A_{k]}
= (\partial_iH^i{}_{[j})A_{k]}.
\end{equation}
Accordingly, we find {\it identically}
\begin{equation}
\Delta\Sigma_{[jk]} + \partial_iS_{jk}{}^i \equiv -\,{\frac {\delta L^{\rm e}}
{\delta A_n}}(\rho_{jk})^m_nA_m.
\end{equation}
Thus we come to the same conclusion that the angular momentum balance 
equation is perfectly satisfied despite the lack of the symmetry of the 
canonical energy-momentum tensor. 

\subsection{Abraham energy-momentum tensor}

In Sec.~\ref{abr}, we described a general procedure for constructing a
symmetric tensor from an arbitrary energy-momentum.  Now we can apply
this scheme to the electromagnetic field, taking the canonical
energy-momentum (\ref{emtE}) as an input. Although we have
demonstrated above that the argument of the ``violation of the angular
momentum conservation" is unsubstantiated for open systems, it seems
reasonable to perform a detailed comparison of the energy-momentum
tensors available on the market.

In this approach, we assume the existence of the timelike vector field
$u^i$, with $u^2 = c^2$. Then substituting (\ref{emtE}) and
(\ref{emtMa}) into the definition (\ref{tAdef}), we construct the
Abraham tensor of the electromagnetic energy-momentum:
\begin{eqnarray}
{\stackrel {\rm A} \Sigma}{}_k{}^i &=& -\,{\frac 12}(H^{ij}F_{kj} + F^{ij}H_{kj})
+ {\frac 14}H^{jl}F_{jl}\,\delta_k^i \nonumber\\
&& +\,{\frac 1{2c^2}}\left[u^iu_l\left(F_{jk}H^{jl} - H_{jk}F^{jl}\right) + 
u_ku^l\left(F^{ji}H_{jl} - H^{ji}F_{jl}\right)\right].\label{emtA}
\end{eqnarray}
In the rest frame, $u^i = \delta^i_0$, we find the Abraham energy flux density
${\stackrel {\rm A} s}{}^a = {\stackrel {\rm A} \Sigma}{}_0{}^a$ and the field 
momentum ${\stackrel {\rm A} p}{}_a = -\,{\stackrel {\rm A} \Sigma}{}_a{}^0$:
\begin{equation}
{\stackrel {\rm A} {\bf p}} = {\frac {\stackrel {\rm A} {\bf s}} {c^2}} = 
{\frac {{\bf E}\times{\bf H}}{c^2}}.\label{psA}
\end{equation}
The Abraham force (\ref{diff2}) then finally reduces to its well known 
rest-frame expression 
\begin{equation}\label{diff3}
{\stackrel {A} {\bf f}}{} = {\bf f} + {\frac {\partial}{\partial t}}\left(
{\bf D}\times {\bf B} - {\frac {{\bf E}\times{\bf H}}{c^2}}\right) + {\frac 12}
{\bf \nabla}\times \left({\bf D}\times{\bf E} + {\bf H}\times{\bf B}\right).
\end{equation}

The last word in deciding which energy-momentum tensor is correct (and
in which sense) belongs certainly to experiment. However, the
theoretical foundation of the Minkowski tensor obviously appears to be
far more solid than that of the Abraham tensor. The former arises as a
canonical energy-momentum tensor in the Lagrange-Noether framework,
whereas the latter one is not derived from first principles.
Nevertheless, as we will demonstrate in the next section, the Abraham
tensor does resurface after we specify the structure and the dynamics
of matter.

\section{Variational model of matter}\label{mat}

Let us now consider the dynamics of the medium. We will model the
matter as an ideal fluid, the elements of which are structureless
particles (i.e., no spin or other internal degrees of freedom are
present). Such a continuous medium (see \cite{taub,schutz,bailyn}, for
example, for the relevant earlier work on the relativistic ideal
fluids) is characterized in the Eulerian approach by the fluid
4-velocity $u^i$, the {\it internal energy} density $\rho$, the {\it
  particle density} $\nu$, the {\it entropy} density $s$, and the {\it
  identity (Lin) coordinate} $X$. Furthermore, we assume that the
motion of a fluid is such that the number of particles is constant and
that the entropy and the identity of the elements is conserved. In
mathematical terms this means that the following constraints are
imposed on the variables:
\begin{eqnarray}
\partial_i(\nu u^i) &=& 0,\label{numbercons}\\
u^i\partial_is &=& 0,\label{entrcons}\\
u^i\partial_iX &=& 0.\label{idcons}
\end{eqnarray}  
Due to the conservation of the entropy only reversible processes are
allowed.  In a variational approach, these constraints are taken into
account by means of Lagrange multipliers. The classical action of the
fluid reads $I^{\rm m} = \int L^{\rm m}\,d^4x$, with the Lagrangian
density
\begin{equation}\label{Lfluid}
L^{\rm m} = -\,\rho(\nu, s) + \Lambda_0(u^iu_i - c^2) - \nu u^i\partial_i
\Lambda_1 + \Lambda_2\,u^i\partial_is + \Lambda_3\,u^i\partial_iX.
\end{equation}
The Lagrange multipliers $\Lambda_1, \Lambda_2, \Lambda_3$ impose the 
constraints (\ref{numbercons})-(\ref{idcons}) on the dynamics of the fluid, 
whereas $\Lambda_0$ accounts for the normalization condition for 
the 4-velocity
\begin{equation}
g_{ij}u^iu^j = c^2.\label{u2}
\end{equation}
For the description of the thermodynamical properties of the fluid, the 
usual thermodynamical law (``Gibbs relation") is used,
\begin{equation}
T\,ds = d(\rho/\nu) + p\,d(1/\nu),\label{thermod}
\end{equation}
where $T$ is the temperature and $p$ the pressure. From this we have 
\begin{eqnarray}
{\frac {\partial\rho}{\partial s}} &=& \nu T,\label{drs}\\
{\frac {\partial\rho}{\partial \nu}} &=& {\frac {\rho + p}\nu}.\label{drnu}
\end{eqnarray}
One can in fact treat the above equations as the definition the 
temperature and the pressure.

\subsection{Equations of motion}

Recalling the general formalism of Sec.~\ref{open}, we describe the 
material system (ideal fluid) by a collective field as $\Phi^A = (u^i, 
\nu, s, X, \Lambda_0, \Lambda_1, \Lambda_2, \Lambda_3)$. The last six
variables characterize only matter, i.e., they enter the matter 
Lagrangian $L^{\rm m}$ but not the Lagrangian of the electromagnetic
field $L^{\rm e}$. The latter depends, in general, on the velocity 
of the medium $u^i$ and on the particle density $\nu$. This is a
manifestation of the fact that both systems (electromagnetic and 
material) are {\it open}. 

The variational derivatives $\delta L^{\rm m}/\delta\Lambda_K = 0$,
$K = 0,1,2,3,$ with respect to the Lagrange multipliers yield the 
constraints (\ref{numbercons})-(\ref{idcons}), (\ref{u2}), whereas 
variation of $L^{\rm m}$ with respect to $s$ and $X$ yields, respectively,
\begin{eqnarray}
{\frac {\delta L^{\rm m}}{\delta s}} &=& \partial_i\left(\Lambda_2 u^i
\right) + {\frac {T\nu} c} = 0,\label{dVds}\\
{\frac {\delta L^{\rm m}}{\delta X}} &=& \partial_i\left(\Lambda_3 u^i
\right) = 0.\label{dVdX}
\end{eqnarray}
In order to derive (\ref{dVds}), we used (\ref{drs}).

It remains to find the variations with respect to $\nu$ and $u^i$. The
direct calculation (use (\ref{drnu})) yields
\begin{eqnarray}
{\frac {\delta L^{\rm m}}{\delta \nu}} &=& -\,{\frac {\rho + p} \nu}
- u^i\partial_i\Lambda_1,\label{dLnu}\\
{\frac {\delta L^{\rm m}}{\delta u^i}} &=& 2\Lambda_0u_i - \nu\partial_i
\Lambda_1 + \Lambda_2\partial_is + \Lambda_3\partial_iX.\label{dLu}
\end{eqnarray}
We cannot put these equations equal to zero because the material system is
open. Contracting (\ref{dLu}) with $u^i$, after using (\ref{dLnu}) and
the constraints (\ref{entrcons}) and (\ref{idcons}), we find the Lagrange 
multiplier:
\begin{equation}
2\Lambda_0 = {\frac 1{c^2}}\left(-\rho - p - \nu{\frac {\delta L^{\rm m}}
{\delta \nu}} + u^i{\frac {\delta L^{\rm m}}{\delta u^i}}\right).
\end{equation}
Substituting this back into (\ref{dLu}), we derive the following useful 
relation
\begin{equation}\label{pi}
- \nu\partial_i\Lambda_1 + \Lambda_2\partial_is + \Lambda_3\partial_iX = 
{\frac {u_i}{c^2}}\left(\rho + p + \nu{\frac {\delta L^{\rm m}}{\delta \nu}}
\right) + {\stackrel \bot P}{}_i^j{\frac {\delta L^{\rm m}}{\delta u^j}}.
\end{equation}
Here, as usual, ${\stackrel \bot P}{}_j^i = \delta_j^i - u^iu_j/c^2$
is the transversal projector. Substituting this into (\ref{Lfluid}),
we see that {\it ``on-shell"} (i.e., if the equations of motion are
fulfilled) the Lagrangian of the medium satisfies
\begin{equation}\label{Lonshell}
L^{\rm m} = p + \nu{\frac {\delta L^{\rm m}}{\delta \nu}}.
\end{equation}

\subsection{Canonical energy-momentum of matter}

Since $L^{\rm m}$ depends only on the derivatives of $s, X$,
and $\Lambda_1$, we have
\begin{equation}
{\frac {\partial L^{\rm m}}{\partial (\partial_i\Phi^A)}}\,\partial_k\Phi^A
= u^i\left(- \nu\partial_k\Lambda_1 + \Lambda_2\partial_ks + \Lambda_3
\partial_kX\right).
\end{equation}
Using (\ref{pi}) and (\ref{Lonshell}), we then straightforwardly construct,
{}from the definition (\ref{emt}), the {\it canonical energy-momentum 
tensor of matter}:
\begin{equation}
{\stackrel {\rm m} \Sigma}{}_k{}^i = u^i{\cal P}_k + \left(-\delta_k^i 
+ {\frac 1{c^2}}u_ku^i\right)p^{\rm eff}.\label{emtF}
\end{equation}
Here we denoted
\begin{eqnarray}
{\cal P}_k &=& {\frac \rho {c^2}}u_k + {\frac {\delta L^{\rm m}}
{\delta u^k}} - {\frac {u_ku^j}{c^2}}{\frac {\delta L^{\rm m}}
{\delta u^j}},\label{Pk}\\
p^{\rm eff} &=& p + \nu{\frac {\delta L^{\rm m}}{\delta \nu}}.\label{peff}
\end{eqnarray}
The physical interpretation of these quantities is clear. The 4-vector 
(\ref{Pk}) is the relativistic 4-momentum density effectively carried by 
the elements of matter. The first term on the right-hand side is the 
usual kinetic momentum determined by the mass (energy density) of the
particles, whereas the two next terms arise from the interaction of the 
medium with the electromagnetic field. The same applies to the second
term of the effective pressure (\ref{peff}) which ``corrects" the usual
hydrodynamical pressure by the term arising due to the fact that the 
material system is open. For the closed system we would have to put
$\delta L^{\rm m}/\delta u^k = 0$ and $\delta L^{\rm m}/\delta \nu =0$,
and then the energy-momentum (\ref{emtF}) would reduce to the standard
expression of the ideal fluid. 

Since the material system is open, the divergence of the energy-momentum
tensor is nontrivial. The energy-momentum balance equation (\ref{dt}) 
now reads
\begin{equation}\label{dtF}
\partial_i\,{\stackrel {\rm m} \Sigma}{}_k{}^i = -\,{\frac {\delta 
L^{\rm m}}{\delta\nu}}\,\partial_k\nu - {\frac {\delta L^{\rm m}}
{\delta u^i}}\,\partial_ku^i.
\end{equation}
As usual, the right-hand side describes the forces that the
electromagnetic field exerts on the matter.

Let us now inspect the angular momentum balance equation. As a first
step, we notice that only the 4-velocity vector field $u^i$ of the
material variable $\Phi^A = (u^i, \nu, s, X, \Lambda_0, \Lambda_1,
\Lambda_2, \Lambda_3)$ has nontrivial transformation properties under
the action of the Lorentz group. The corresponding generators read
$(\rho^j{}_k)^m_n = \delta^{[j}_n\delta^m_{k]}$.  However, since
$\partial L^{\rm m}/\partial(\partial_ju^i) = 0$, the spin density of
the medium vanishes, $S_{jk}{}^i = 0$.

The canonical energy-momentum of matter (\ref{emtF}) is not symmetric, 
\begin{equation}
{\stackrel {\rm m} \Sigma}{}_{[jk]} = {\cal P}_{[j}u_{k]} = {\frac {\delta 
L^{\rm m}}{\delta u^{[j}}}\,u_{k]}. 
\end{equation}
However, the right-hand side of (\ref{dS}) is now
\begin{equation}
-\,{\frac {\delta L^{\rm m}}{\delta u^m}}\,(\rho_{jk})^m_n u^n = 
-\,{\frac {\delta L^{\rm m}}{\delta u^{[k}}}\,u_{j]}. 
\end{equation}
We thus verify that the angular momentum balance equation (\ref{dS}) is
satisfied 
\begin{equation}
{\stackrel {\rm m} \Sigma}{}_{[jk]} \equiv -\,{\frac {\delta L^{\rm m}}
{\delta u^m}}\,(\rho_{jk})^m_n u^n\label{dS1}
\end{equation}
{\it identically} for the asymmetric canonical energy-momentum (\ref{emtF})
of the medium. 

\section{Coupled system of the electromagnetic field and matter}

We finally can complete the picture by combining the two pieces which 
we studied separately above: the electromagnetic system of Sec.~\ref{maxwell} 
and the material system of Sec.~\ref{mat}. In order to do this, we need one
important additional input. Namely, we need to specify how exactly the two 
systems interact. In the most general form, the information about this 
interaction is encoded in the constitutive tensor density $\chi^{ijkl}$ 
in the electromagnetic Lagrangian (\ref{Le}). 

The model of the fluid, which we studied in the previous section, obviously
describes the polarizable/magnetizable medium with isotropic electric and
magnetic properties. Such matter is characterized by the two scalar 
quantities: the permittivity $\varepsilon$ and the permeability $\mu$. 
In order to take into account the possible {\it electrostriction and
magnetostriction} effects, we allow for these quantities to be the 
functions of the particle density, 
\begin{equation}
\varepsilon = \varepsilon(\nu),\qquad \mu = \mu(\nu).\label{epsmu}
\end{equation}

We thus assume that the medium is intrinsically {\it isotropic}, and
the possible {\it anisotropic} effects may arise only from the
nontrivial motion of the medium, i.e., they are induced by the
velocity of the fluid.

\subsection{Constitutive relation and the electromagnetic Lagrangian}
\label{maxmodel}

The constitutive relation for an arbitrarily moving isotropic medium is 
well known. It is given by the famous Minkowski equations, see \cite{HO03}, 
Sec.E.4.2, eqs.(E.4.28)-(E.4.29) and (E.4.25)-(E.4.26). The constitutive 
tensor for this case is given by the formula that is very close to 
(\ref{chi0}), namely, 
\begin{equation}
\chi{}^{ijkl} = \lambda\sqrt{-g_{\rm opt}}\left(g_{\rm opt}^{ik}g_{\rm opt}^{jl} 
- g_{\rm opt}^{il}g_{\rm opt}^{jk}\right),\label{chi1}
\end{equation}
where $\lambda = \sqrt{\varepsilon\varepsilon_0/\mu\mu_0}$ and the so-called 
{\it optical metric} was first introduced by Gordon \cite{gordon0}:
\begin{equation}
g_{\rm opt}^{ij} = g^{ij} - {\frac {1 - n^2}{c^2}}\,u^i\,u^j.\label{gopt}
\end{equation}
Here $n = \sqrt{\varepsilon\mu}$ is the refraction coefficient of the medium.

We straightforwardly find $\sqrt{-g_{\rm opt}} = c/n$, and thus the Lagrangian
(\ref{Le}) of the electromagnetic field finally reads
\begin{equation}
L^{\rm e} = -\,{\frac 1{4\mu_0\mu}}\,g_{\rm opt}^{ij}\,g_{\rm opt}^{kl}
\,F_{ik}F_{jl} = -\,{\frac 1{4\mu_0\mu}}\left[F_{ij}F^{ij} + 2{\frac {n^2 -1}
{c^2}}\,F_{ik}u^kF^{il}u_l\right].\label{Lem}
\end{equation}
The components of the electromagnetic field $F_{ij}$ are given with respect
to the laboratory system. In order to get some further insight into the 
formulas above, we notice that the electric and magnetic fields in the 
comoving system (in which the medium is momentarily at rest) can be described
by the 4-vectors 
\begin{equation}
{\cal E}_i := F_{ik}u^k,\qquad {\cal B}^i := {\frac {1}{2c}}
\eta^{ijkl}F_{jk}u_l.\label{EBvec4}
\end{equation}
The field strength is uniquely reconstructed from these vectors as
\begin{equation}
F_{ij} = {\frac 1 {c^2}}\left({\cal E}_iu_j - {\cal E}_ju_i + c\eta_{ijkl}
u^k{\cal B}^l\right).\label{FEB}
\end{equation}
The electromagnetic Lagrangian (\ref{Lem}) looks remarkably simple in these
variables:
\begin{equation}
L^{\rm e} = -\,{\frac 12}\left(\varepsilon\varepsilon_0{\cal E}^2 -
{\frac {{\cal B}^2}{\mu\mu_0}}\right),\label{Lem2}
\end{equation}
with the obvious abbreviations ${\cal E}^2 = {\cal E}_i{\cal E}^i$ and 
${\cal B}^2 = {\cal B}_i{\cal B}^i$. Both vectors are by construction 
orthogonal to the 4-velocity, ${\cal E}_iu^i = 0$ and ${\cal B}^iu_i = 0$.
Hence, ${\cal E}^2 \leq 0$ and ${\cal B}^2 \leq 0$.

In Sec.~\ref{EM} we treated the whole constitutive tensor density 
$\chi^{ijkl}$ as the material variable. Now, after specifying the model
of the medium, we have something better. The fluid under consideration is 
described by the variables $u^i$ and $\nu$, and the constitutive tensor
now becomes a known function $\chi^{ijkl}=\chi^{ijkl}(u^m,\nu)$, given 
by the equation (\ref{chi1}). The corresponding variational derivatives
of the electromagnetic Lagrangian with respect to the material variables
are easily computed:
\begin{eqnarray}
{\frac {\delta L^{\rm e}}{\delta u^i}} &=& -\,{\frac {n^2 -1}{\mu\mu_0c^2}}
F_{ki}F^{kl}u_l,\label{dLeu}\\
{\frac {\delta L^{\rm e}}{\delta \nu}} &=& -\,{\frac 12}\left(\varepsilon_0
{\frac {\partial\varepsilon}{\partial\nu}}\,{\cal E}^2 + {\frac 1{\mu_0
\mu^2}}{\frac {\partial\mu}{\partial\nu}}\,{\cal B}^2\right).\label{dLen}
\end{eqnarray}
We used (\ref{Lem}) to derive (\ref{dLeu}), however, it is much simpler
to use a different (equivalent) form of the Lagrangian (\ref{Lem2}) to
obtain (\ref{dLen}).

\subsection{Canonical energy-momentum tensor}

After all these preliminaries, we are now in a position to find the explicit 
form of the energy-momentum tensors and to analyse the corresponding balance
equations. 

The electromagnetic excitation (\ref{Hij}) for the Lagrangian (\ref{Lem}) is
\begin{equation}
H^{ij} = {\frac 1{\mu_0\mu}}\,g_{\rm opt}^{ik}\,g_{\rm opt}^{jl}
\,F_{kl} = {\frac 1{\mu_0\mu}}\left[F^{ij} + {\frac {n^2 - 1}{c^2}}
\left(F^{ik}u_ku^j - F^{jk}u_ku^i\right)\right].\label{Hij0}
\end{equation}
Substituting this into (\ref{emtE}), we find the explicit form of the
canonical (Minkowski) energy-momentum of the electromagnetic field:
\begin{eqnarray}
{\stackrel {\rm e} \Sigma}{}_k{}^i &=& {\frac 1{\mu\mu_0}}\bigg[
-\,F^{ij}F_{kj} + {\frac 1{4}}F^{jl}F_{jl}\,\delta_k^i\nonumber\\ 
\label{emtEF} 
&& +\,{\frac {n^2 - 1}{c^2}}\left(F_{kn}F^{nl}u_lu^i - F_{kl}u^l
F^{in}u_n + {\frac {1}{2}}F_{nl}u^lF^{nj}u_j\delta_k^i\right)\bigg].
\end{eqnarray}
This tensor is not symmetric, and its antisymmetric part (\ref{emtMa})
now reads explicitly
\begin{equation}\label{emtMskew}
{\stackrel {\rm e} \Sigma}{}_{[jk]} = -\,{\frac {n^2 - 1}{\mu\mu_0c^2}}
\,u_{[j}F_{k]i}F^{il}u_l.
\end{equation}

\subsection{Balance equations of the energy-momentum and angular momentum}

As we know, the Minkowski tensor (\ref{emtEF}) is not conserved, and the 
energy-momentum balance equation (\ref{dtM}) contains a nontrivial force 
on the right-hand side, $X_k = {\frac 18}\,F_{ij}F_{mn}\,\partial_k
\chi^{ijmn}$. Since $\chi^{ijkl}=\chi^{ijkl}(u^m,\nu)$, we have 
\begin{equation}
\partial_k\chi^{ijmn} = {\frac {\partial\chi^{ijmn}}{\partial \nu}}
\,\partial_k\nu + {\frac {\partial\chi^{ijmn}}{\partial u^m}}
\,\partial_ku^m.
\end{equation}
Then the force density can be easily calculated,
\begin{eqnarray}
X_k &=& {\frac {\partial}{\partial \nu}}\left({\frac 18}\,\chi^{ijkl}
\,F_{ij}F_{kl}\right)\partial_k\nu + {\frac {n^2 -1}{\mu\mu_0c^2}}
F^{ij}u_jF_{im}\,\partial_ku^m\nonumber\\
&=& -\,{\frac {\delta L^{\rm e}}{\delta \nu}}\,\partial_k\nu
- {\frac {\delta L^{\rm e}}{\delta u^i}}\,\partial_ku^i,\label{XkF}
\end{eqnarray}
where we used (\ref{Le}) and (\ref{dLeu}).

Furthermore, for the explicit dependence of the constitutive tensor 
(\ref{chi1}) on $u^i$ and for the Lorentz generators $(\rho^j{}_k)^m_n 
= \delta^{[j}_n\delta^m_{k]}$ of the 4-velocity vector field, we prove
the identity
\begin{equation}
-\,{\frac {\delta L^{\rm e}}{\delta \chi^{mnpq}}}(\rho_{jk}
)^{mnpq}_{m'n'p'q'}\chi^{m'n'p'q'} \equiv -\,{\frac {\delta L^{\rm e}}
{\delta u^m}}(\rho_{jk})^m_n\,u^n. 
\end{equation}
Making use of the variational derivative (\ref{dLeu}), we have 
\begin{equation}
-\,{\frac {\delta L^{\rm e}}{\delta u^m}}(\rho_{jk})^m_n\,u^n =
-\,{\frac {n^2 - 1}{\mu\mu_0c^2}}\,u_{[j}F_{k]n}F^{nl}u_l. 
\end{equation}
Comparing with (\ref{emtMskew}), we thus verify that the balance 
equation of the angular momentum
\begin{equation}\label{dS2}
{\stackrel {\rm e} \Sigma}{}_{[jk]} \equiv -\,{\frac {\delta L^{\rm e}}
{\delta u^m}}(\rho_{jk})^m_n\,u^n
\end{equation}
is identically satisfied in this case, as before.

\subsection{Total energy-momentum tensor of the coupled system}

Now, the crucial step is to recall that the total system $L^{\rm tot}
= L^{\rm e} + L^{\rm m}$ of the coupled electromagnetic field and 
medium is {\it closed}. Thus, 
\begin{equation}
{\frac {\delta L^{\rm e}}{\delta \nu}} + {\frac {\delta L^{\rm m}}
{\delta \nu}} = 0,\qquad {\frac {\delta L^{\rm e}}{\delta u^i}} +
{\frac {\delta L^{\rm m}}{\delta u^i}} = 0.\label{eqm}
\end{equation}
By combining (\ref{XkF}) with (\ref{dtF}), we then finally express the 
force that acts on the electromagnetic field in terms of the material
variables,
\begin{equation}
X_k = -\,\partial_i\,{\stackrel {\rm m} \Sigma}{}_k{}^i.\label{Xk0}
\end{equation}

Substituting this into (\ref{dtM}), and rearranging the terms, we find
the true {\it conservation law} of the total energy-momentum of the
closed system (field + medium):
\begin{equation}
\partial_i\left({\stackrel {\rm e} \Sigma}{}_k{}^i + {\stackrel {\rm m} 
\Sigma}{}_k{}^i\right) = 0.\label{dtTot}
\end{equation}
Note that we assumed that the medium is {\it neutral}, hence $J^i =0$.

Using the equations of motion (\ref{eqm}), we can find the explicit form
of the canonical energy-momentum of medium, too. In particular, combining
(\ref{eqm}) with (\ref{dLeu}) and (\ref{dLen}), we have
\begin{eqnarray}
{\cal P}_k &=& {\frac \rho {c^2}}u_k + {\frac {n^2 - 1}{\mu\mu_0c^2}}\left(
F_{ik}F^{il}u_l - {\frac {u_k}{c^2}}\,F_{jn}u^nF^{jl}u_l\right),\label{PkF}\\
p^{\rm eff} &=& p + {\frac 12}\left(\varepsilon_0{\frac {\partial\varepsilon}
{\partial\nu}}\,{\cal E}^2 + {\frac 1{\mu_0\mu^2}}{\frac {\partial\mu}
{\partial\nu}}\,{\cal B}^2\right).\label{presseff}
\end{eqnarray}
Remarkably, the effective pressure describes the electro- and magnetostriction
effects. Substituting (\ref{PkF}) into (\ref{emtF}), we obtain the final 
expression
\begin{eqnarray}
{\stackrel {\rm m} \Sigma}{}_k{}^i &=& {\frac \rho {c^2}}u_ku^i + \left(
-\delta_k^i + {\frac 1{c^2}}u_ku^i\right)p^{\rm eff}\nonumber\\
&& +\,{\frac {n^2 - 1}{\mu\mu_0c^2}}\left( u^iF_{nk}F^{nl}u_l 
- {\frac {u_ku^i}{c^2}}\,F_{jn}u^nF^{jl}u_l\right).\label{emtF0}
\end{eqnarray}

The total canonical energy-momentum tensor is the sum of the electromagnetic
(Minkowski) tensor (\ref{emtEF}) and of the energy-momentum of the fluid
(\ref{emtF0}). After some simple algebra we find
\begin{eqnarray}
{\stackrel {\rm e} \Sigma}{}_k{}^i + {\stackrel {\rm m} \Sigma}{}_k{}^i &=& 
{\frac \rho {c^2}}u_ku^i + \left(-\delta_k^i + {\frac 1{c^2}}u_ku^i\right)
p^{\rm eff} + {\frac 1{\mu\mu_0}}\bigg[ -\,F^{ij}F_{kj} + {\frac 1{4}}
F^{jl}F_{jl}\,\delta_k^i\nonumber\\ 
&& +\,{\frac {n^2 - 1}{c^2}}\left(- F_{kl}u^lF^{in}u_n - {\frac {u_ku^i}
{c^2}}\,F_{jn}u^nF^{jl}u_l + {\frac {1}{2}}F_{nl}u^lF^{nj}u_j\delta_k^i
\right)\bigg].\label{emt0}
\end{eqnarray}

It is worthwhile to note that whereas the Minkowski electromagnetic
tensor and the energy-momentum of the fluid are both asymmetric, the
total canonical energy-momentum is explicitly symmetric. This is in
agreement with the fact that the total system is closed. Indeed,
combining the balance equations (\ref{dS1}) and (\ref{dS2}) of the
angular momenta of the electromagnetic and material systems, we find
\begin{equation}\label{dStot}
{\stackrel {\rm e} \Sigma}{}_{[jk]} + {\stackrel {\rm e} \Sigma}{}_{[jk]}
= -\left({\frac {\delta L^{\rm e}}{\delta u^m}} + {\frac {\delta L^{\rm m}}
{\delta u^m}}\right)(\rho_{jk})^m_n\,u^n = 0. 
\end{equation}
To derive the last equality, we used the field equations (\ref{eqm}). 

\subsection{Abraham energy-momentum tensor}

For completeness, let us compute the Abraham energy-momentum. In accordance
with the definition (\ref{tAdef}), we find the projections of the 
antisymmetric part ${\stackrel {\rm e} a}{}_{jk} = {\stackrel {\rm e} 
\Sigma}{}_{[jk]}$ of the Minkowski tensor (\ref{emtMskew}):
\begin{equation}\label{tMpro}
{\stackrel \bot P}{}_k^j\,{\stackrel {\rm e} a}{}_j{}^i = {\stackrel {||} 
P}{}_j^i\,{\stackrel {\rm e} a}{}_k{}^j = {\frac {n^2 - 1}{2\mu\mu_0c^2}}
\,u^i\left(F_{kn} - {\frac {u_ku^j}{c^2}}\,F_{jn}\right)F^{nl}u_l.
\end{equation}
The resulting electromagnetic Abraham energy-momentum tensor reads
\begin{eqnarray}
{\stackrel {\rm A} \Sigma}{}_k{}^i &=& {\stackrel {\rm e} \Sigma}{}_k{}^i
- {\stackrel \bot P}{}_k^j\,{\stackrel {\rm e} a}{}_j{}^i - {\stackrel 
{||} P}{}_j^i\,{\stackrel {\rm e} a}{}_k{}^j = {\frac 1{\mu\mu_0}}\bigg[ 
-\,F^{ij}F_{kj} + {\frac 1{4}}F^{jl}F_{jl}\,\delta_k^i\nonumber\\ 
&& +\,{\frac {n^2 - 1}{c^2}}\left(- F_{kl}u^lF^{in}u_n - {\frac {u_ku^i}
{c^2}}\,F_{jn}u^nF^{jl}u_l + {\frac {1}{2}}F_{nl}u^lF^{nj}u_j\delta_k^i
\right)\bigg].\label{emtA0}
\end{eqnarray}
The same expression is obtained, after some algebra, if we insert the 
electromagnetic excitation tensor (\ref{Hij0}) directly in the general 
definition (\ref{emtA}).

Comparing the formula (\ref{emtA0}) with (\ref{emt0}), we observe that all 
the energy-momentum tensors are related via the interesting equation
\begin{equation}
{\stackrel {\rm e} \Sigma}{}_k{}^i + {\stackrel {\rm m} \Sigma}{}_k{}^i
= {\frac \rho {c^2}}u_ku^i + \left(-\delta_k^i + {\frac 1{c^2}}u_ku^i\right)
p^{\rm eff} + {\stackrel {\rm A} \Sigma}{}_k{}^i.\label{Ttot}
\end{equation}
So to say, the Abraham tensor ``absorbed" all the terms which explicitly 
contained the electromagnetic field, except for the electrostriction  and
magnetostriction terms, both from the Minkowski canonical tensor of the 
electromagnetic field and from the canonical energy-momentum of the matter. 

\section{Why Abraham?}

The fact that the Abraham tensor resurfaced at the end of our
derivations requires an additional analysis. Whereas the Minkowski
tensor has a solid standing as a canonical energy-momentum deeply
rooted in the Lagrange-Noether formalism, the Abraham tensor appears a
rather artificial construct, as we saw in Sec.~\ref{abr}. So, why is
it recovered in the equation (\ref{Ttot})?  Is this a coincidence, a
specific feature of the models that we used for the description of the
electromagnetic and material systems, or something more fundamental?
Here we demonstrate that the latter is true, in a certain sense.

\subsection{A useful mathematical fact}\label{fact}

At first, we prove the following technical point. Let $\alpha_{ij} = 
-\,\alpha_{ji}$ be an antisymmetric tensor and $\beta_k$ a 4-vector. 
Then $\alpha_{ij}$ satisfies the algebraic equation 
\begin{equation}
{\stackrel \bot P}{}_k^j\,\alpha_j{}^i + {\stackrel {||} P}{}_j^i
\,\alpha_k{}^j = u^i\,{\stackrel \bot P}{}_k^j\,\beta_j\label{ab1}
\end{equation}
if and only if this tensor is constructed from $\beta_k$ and the velocity 
$u^i$ as
\begin{equation}
\alpha_{ij} = \beta_{[i}\,u_{j]}.\label{ab2}
\end{equation}

The proof is straightforward. Let us assume that (\ref{ab2}) is true. Then
\begin{eqnarray}
{\stackrel \bot P}{}_k^j\,\alpha_j{}^i &=& \left(\delta_k^j - {\frac {u_ku^j}
{c^2}}\right){\frac 12}\left(\beta_ju^i - \beta^iu_j\right) = {\frac {u^i}2}
\,{\stackrel \bot P}{}_k^j\,\beta_j,\\
{\stackrel {||} P}{}_j^i\,\alpha_k{}^j &=& {\frac {u_ju^i}{c^2}}\,{\frac 12}
\left(\beta_ku^j - \beta^ju_k\right) = {\frac {u^i}2}\,{\stackrel \bot P}
{}_k^j\,\beta_j.
\end{eqnarray}
Hence, (\ref{ab1}) is fulfilled. 

Conversely, suppose (\ref{ab1}) is true. Contracting this equation 
with $u^i$, we find
\begin{equation}
{\stackrel \bot P}{}_k^j\,u_i\alpha_j{}^i + u_i{\stackrel {||} P}{}_j^i
\,\alpha_k{}^j = 2\,u_i\alpha_k{}^i = c^2\,{\stackrel \bot P}{}_k^j
\,\beta_j\quad\Longrightarrow\quad u^j\alpha_j{}^i = -\,{\frac {c^2}2}
\,{\stackrel \bot P}{}_j^i\,\beta^j.\label{ab3}
\end{equation}
On the other hand, from (\ref{ab1}) we have, making use of (\ref{ab3}):
\begin{equation}
{\stackrel \bot P}{}_k^j\,\alpha_j{}^i = -\,{\stackrel {||} P}{}_j^i
\,\alpha_k{}^j + u^i\,{\stackrel \bot P}{}_k^j\,\beta_j = {\frac 12}
\,u^i{\stackrel \bot P}{}_k^j\,\beta_j.\label{ab4}
\end{equation}
This finally yields, again using (\ref{ab3}), 
\begin{equation}
\alpha_k{}^i = \left({\stackrel \bot P}{}_k^j + {\stackrel {||} P}
{}_k^j\right)\alpha_j{}^i = {\frac 12}\left(u^i{\stackrel \bot P}{}_k^j
\,\beta_j - u_k{\stackrel \bot P}{}_j^i\,\beta^j\right) = {\frac 12}
\left(u^i\beta_k - u_k\beta^i\right).
\end{equation}
Thus, the statement is proved.

\subsection{Crucial role of velocity}

We now extend the electrodynamical model of Sec.~\ref{maxmodel} by
generalizing the constitutive law (\ref{chi1}). Namely, we assume that
the constitutive tensor $\chi^{ijkl}$ is not a particular function
(\ref{chi1}) of the 4-velocity $u^i$ and of the particle density
$\nu$, but it rather is some general function
\begin{equation}
\chi^{ijkl} = \chi^{ijkl}(u^m, \nu, \psi^{\Omega}).\label{chi2}
\end{equation}
Here the collective variable $\psi^{\Omega}$ denotes all possible 
parameters that describe the state of the polarizable/magnetizable 
medium. In the macroscopic approach, $\psi^{\Omega}$ includes the
most general matrices of dielectric, magnetic, and magnetoelectric
susceptibilities. An important assumption is that these parameters
are {\it not changed} under the Lorentz transformations relating 
different reference systems. 

More precisely, we thus assume that the 4-velocity of the medium $u^i$ 
is {\it the only (Lorentz) covariant object} that enters the 
electrodynamical constitutive relation (\ref{chi2}), and the other 
material variables $(\nu, \psi^{\Omega})$ are all scalars under the 
Lorentz transformations. Technically, this is realized by assuming 
that the parameters $\psi^{\Omega}$ take their genuine (or intrinsic) 
values in the comoving reference system with respect to which the 
medium is momentarily at rest. In other reference systems, these 
intrinsic values of $\psi^{\Omega}$ remain the same, whereas in 
every reference system the form of the constitutive matrices is 
determined {\it only by the motion of the medium}, i.e., in technical 
terms, by the velocity $u^i$. 

This assumption has the far-reaching consequences. Since
the velocity $u^i$ is the only covariant argument (with the Lorentz 
generators $(\rho^j{}_k)^m_n = \delta^{[j}_n\delta^m_{k]}$) of the 
constitutive tensor $\chi^{ijkl}(u^m, \nu, \psi^{\Omega})$, we have 
\begin{equation}
{\frac {\delta L^{\rm e}}{\delta \chi^{mnpq}}}(\rho_{jk})^{mnpq}_{m'n'p'q'}
\chi^{m'n'p'q'} = {\frac {\delta L^{\rm e}}{\delta \chi^{mnpq}}}{\frac 
{\partial\chi^{mnpq}}{\partial u^m}}(\rho_{jk})^m_n\,u^n = 
{\frac {\delta L^{\rm e}}{\delta u^m}}\,(\rho_{jk})^m_n u^n = 
{\frac {\delta L^{\rm e}}{\delta u^{[k}}}\,u_{j]}.
\end{equation}
[Hint: denoting $\omega^{jk}=-\omega^{kj}$ the parameters of the Lorentz 
group, $\delta_{\omega}\chi^{mnpq} = \omega^{jk}(\rho_{jk})^{mnpq}_{m'n'p'q'}
\chi^{m'n'p'q'}$ gives the infinitesimal Lorentz transformation; 
using for differentiation the chain rule completes the proof.]
As a result, we notice that the angular momentum balance equation
(\ref{angbal}) is of the form (\ref{ab2}) where
\begin{equation}
\alpha_{ij} = {\stackrel {\rm e} \Sigma}{}_{[ij]},\qquad
\beta_k = {\frac {\delta L^{\rm e}}{\delta u^k}}.
\end{equation}
Accordingly, combining (\ref{ab1}) with the definition of the Abraham
tensor (\ref{tAdef}), we find the relation
\begin{equation}\label{MArel}
{\stackrel {\rm e} \Sigma}{}_k{}^i = {\stackrel {\rm A} \Sigma}{}_k{}^i
+ u^i\,{\stackrel \bot P}{}_k^j\,{\frac {\delta L^{\rm e}}{\delta u^j}}.
\end{equation}
Adding the canonical energy-momentum (\ref{emtF}), (\ref{Pk}), we then
finally come to the equation (\ref{Ttot}).

This analysis demonstrates that the relation (\ref{Ttot}) between the 
canonical energy-momentum tensors and the Abraham energy-momentum is
not occasional. It is valid not only for the isotropic constitutive 
relation (\ref{chi1}) of the specific model which we studied in 
Sec.~\ref{maxmodel}, but for the general constitutive relation 
(\ref{chi2}) as well. Sec.~\ref{fact} provides an explanation why
the abrahamization prescription (\ref{tAdef}) is successful, after all.

The key to the Abraham tensor is in the 4-velocity of the medium:
when $u^i$ is the only (Lorentz) covariant variable that enters the
constitutive relation (and only then), the projections of the Minkowski 
tensor satisfy (\ref{ab1}) that yields (\ref{MArel}). This explains why
the Abraham tensor turns out to be relevant for the discussion of the
energy and momentum of moving media, especially of the isotropic ones.
However, the angular momentum balance equation no longer has the form
(\ref{ab2}) if the constitutive tensor depends on additional covariant
material variables, and the final nice result (\ref{MArel}) is invalid
then. The relevance of the Abraham construction is thus limited. 

\section{Discussion and conclusion}

The complete century-long history of the discussion of the
electromagnetic energy and momentum in moving media is still to be
written. In this short paper, we presented a consistent viewpoint on
this problem within the framework of the Lagrange-Noether approach.
The canonical energy-momentum tensor appears then as a fundamental
structure, together with the relevant balance equations of energy,
momentum, and angular momentum.

By a detailed derivation, we explicitly demonstrate that the angular
momentum balance equation is always perfectly satisfied for the
canonical tensor of the electromagnetic system when one takes care of
the fact that it is an open system. Even in absence of the intrinsic
(spin) degrees of freedom, the canonical energy-momentum tensor of an
open system does not need to be symmetric, contrary to a widely spread
misunderstanding.

The Minkowski energy-momentum tensor (\ref{emtE}) arises as a
canonical structure in macroscopic Max\-wellian electrodynamics, which
stresses its fundamental physical nature. The balance equation of an
open electromagnetic system (\ref{dtM}) contains nontrivial force
terms. Without specifying a model for the medium, we demonstrate that
the $X_k$ force (\ref{X1}) is related to the bound charges and
currents, thus giving rise to the balance equation (\ref{dT}) of the
total canonical energy-momentum tensor (\ref{emtT}). Its validity for
the analysis of the crucial experiments was demonstrated in
\cite{emt}.

Alternatively, the $X_k$ force (\ref{X1}) can be reconstructed in
terms of the material variables when a model of the medium is
specified. We show how this works in the framework of the variational
approach to the relativistic ideal fluid. Here again the canonical
energy-momentum (\ref{emtF}) appears as a fundamental object. The
interaction of matter with the electromagnetic field induces an
additional field-dependent term in the material momentum (\ref{Pk})
superimposed with the usual hydrodynamical structures. The pressure
(\ref{peff}) picks up the electrostriction and magnetostriction
contributions.

In contrast to the canonical structures, the Abraham energy-momentum
tensor does not arise from first principles. We give a formal
prescription that allows one to construct a symmetric tensor
(\ref{tAdef}) from any given energy-momentum, provided a timelike
vector field $u^i$ is available. Unlike the other symmetrization
procedures (such as the Belinfante-Rosenfeld one, for example), the
abrahamization does not preserve the balance equations, it rather
introduces the additional forces in (\ref{diff1}), (\ref{diff2}). When
applied to the Minkowski tensor, this prescription yields the famous
Abraham electromagnetic energy-momentum (\ref{emtA}).

Nevertheless, despite its ad hoc definition, the Abraham tensor 
turns out to be relevant since it resurfaces at the end from the sum of the 
two canonical energy-momenta of the electromagnetic and material systems in 
the eq.\ (\ref{Ttot}). It thus appears as a certain ``hybrid" construct that 
absorbs the explicit field-dependent terms from the canonical energy-momenta
of the electromagnetic and material systems. This curious fact explains why
the Abraham tensor was for such a long time seriously competing with the 
Minkowski tensor in the analyses of the numerous experiments \cite{brevik}.

This result is not confined to the specific model with the
constitutive relation (\ref{chi1}), but holds in general for the class
of constitutive tensor densities (\ref{chi2}) that depend arbitrarily
on the material variables, provided the 4-velocity $u^i$ of the medium
is the only Lorentz-covariant object. This apparently confirms the
recent observations of Dereli, Gratus and Tucker
\cite{Dereli1,Dereli2} that the Abraham energy-momentum arises as the
metrical energy-momentum tensor for media with a general
constitutive relation. If we assume that the metric of spacetime
is not the flat Minkowski one but is a function of coordinates, the
metric energy-momentum tensor density is defined by the variational
derivative
\begin{equation}
\sigma_{ij} := 2c\,{\frac {\delta\left(L^{\rm e} 
+ L^{\rm m}\right)}{\delta g^{ij}}}.\label{Tmet}
\end{equation}
Indeed, a (long but straightforward) computation then shows that 
$\sigma_k{}^i = {\stackrel {\rm e} \Sigma}{}_k{}^i + {\stackrel {\rm m} 
\Sigma}{}_k{}^i$ for the model (\ref{Le}), (\ref{chi1}) and (\ref{Lfluid}) 
of the isotropic moving medium above. There is a difference, strictly 
speaking, between our results since neither the Lagrangian of the fluid 
nor the hydrodynamical part of the energy and momentum is discussed in 
\cite{Dereli1,Dereli2}.

It is worthwhile to stress that the model of the medium above is by no
means a general one. Among its most serious limitations is that the
elements of matter are assumed to be spinless. The extension of the
model to the case of the nontrivial spin (along the lines similar to
those of \cite{Israel1,Israel2,Israel3}) can bring new insight into
the subject, especially with regard to the effects of magnetism.

\begin{acknowledgement}
  My deep thanks go to Friedrich Hehl, with whom most of the results
  described above were obtained, for the constant discussions, support
  and advice. I am also grateful to Bahram Mashhoon for asking
  interesting and important questions. At the earlier stage, this work
  was supported partly by the Deutsche Forschungsgemeinschaft (Bonn)
  with the grant HE 528/21-1.
\end{acknowledgement}


\end{document}